\documentclass{article}
\usepackage[cmex10]{amsmath}
\usepackage{amsfonts}
\usepackage{array,color}
\usepackage{booktabs}
\usepackage{graphicx}
\usepackage{epstopdf}
\usepackage{cite}
\usepackage{algorithm}
\usepackage{algorithmic}
\usepackage{mdwmath}
\usepackage{array}

\usepackage{blindtext}
\usepackage[utf8]{inputenc}
\author{Di Zhang, Takuro Sato\\
Globle Information and Telecommunication Studies, \\
Waseda University, Tokyo, Japan, 169--8050\\
Email: di\_zhang@fuji.waseda.jp
\and
Zhenyu Zhou\\
State Key Laboratory of Alternate Electrical Power System\\
 with Renewable Energy Sources,\\
School of Electrical and Electronic Engineering, \\
North China Electric Power University, Beijing, China, 102206\\
Email: zhenyu\_zhou2008@gmail.com
}
\date{}
\begin{document}

\title{An Energy Efficiency policy for Communications with C-RAN, ICN and Transition Smooth}

\maketitle
  
\begin{abstract}
Towards next generation communications, Energy Efficiency (EE) attracts lots of attentions nowadays. Some innovative techniques have been proposed in prior literatures, especially the sleep mechanism of base station (BS). Yet how to sleep and when to sleep are still vague concepts. Another, most of the studies focus on the cellular section or core networks separately while integral and comprehensive version is neglected in prior literatures. In this paper,the integral optimization structure is studied based on cloud radio network (C-RAN) and information centric network (ICN) that raised latest combined with the sleep mode. The original C-RAN and ICN structures are amended in terms of reality application of sleep techniques.  While adopting the sleep techniques both in core and cellular, apart from previous works, a transition smooth method that solve the current surge problems which is ignored before is further proposed. Based on the new method, it will be much more feasible to adopt the sleep techniques by knowing the appropriate occasion for transition between sleep and idle mode. Comprehensive computer based simulation results demonstrate that this integer proposal achieves better EE feature with negligible impact on quality of service (QoS) of user equipments (UEs).

\end{abstract}

%

\section{Introduction}

According to International Telecommunications Union (ITU), information communication industry (ICT) consume 10\% energy world-wide nowadays while transmitted data volume still increasing approximately with a speed of 10 times every 5 years\cite{1}. Because of the growing energy demands and costs caused by higher speed of transmission, in next generation communications, more attention should be paid on energy efficiency (EE). Thus, some organizations are calling for green cellular communications, for instance, 5GNow, EP-7, etc.\cite{2}. And EE policy also attracts operators' attentions. For instance, orange claims that energy consumption of per user should be reduced by 20\%  comparing with 2006.
In cellular section, it is estimated that nearly 57\% power consumption coming from BS itself. Thanks to the author in\cite{5}, cellular zooming policy is adopted with sleeping mode for BS while transmission. Once the policy is proposed, great deal of attentions are attracted with plenty of further studies based on this. For instance, in\cite{6}, the cooperative transmission of downlink communications with BS and antenna switching off mechanism is proposed; in\cite{7}, the author further extends the study with radio frequency (RF) chain sleeping for multiuser MIMO system; in\cite{8} and our prior work in\cite{9}, the antenna selection mechanisms based on cellular zooming with sleeping mode are studied for massive MIMO systems. But another significant issue comes from China Mobile\cite{11} that ignored by all of those works, which acclaims that nearly30\% power consumption of BS consuming comes from air condition and other facilitate equipments of the machine room. Therefore, more power can be saved by better architecture planning, such as the proposal of cloud radio access network (C-RAN) that raised by China Mobile. But to the best knowledge of us, even the C-RAN proposal itself is still some kind of theoretical concept rather than a realistic prototyping model for latest raise and fewer studies nowadays.
\par
Things in core networks seem familiar, some sleep mode\cite{16} and network coding\cite{15} mechanisms also have been proposed towards energy efficiency. But most of them based on existing network model while seldom attention has been paid on the new technology with ICN, which is of better feature towards reliving the load, better security and EE and proved to be promising technology for future. And to our best knowledge, even less notice has been paid on the coexisting architecture of both although we should focus on with the integrity sight.

\par
Another, although the sleep strategy has been widely studied both in core and cellular network. But while talking about the reality, how to sleep and when to sleep are still an obscure issue, say, as equipments typically associated with uninterrupted power supply (UPS) system, simply turn on or turn off UPS system cannot be a good choice especially within the massive MIMO circumstance; as frequent in-out performance of UEs in cellular, an mechanism is needed for transition between sleep and working mode of antennas; as electron devices (capacitor, inductor, etc.) have some specific working lives and can be significantly destroyed by current pulse during power on-off periods, protection method and trade off mechanism between energy saving of cellular system and working lives of electron devices are greatly needed.

\par
Based on all of these, in this paper, we raise a new answer the question listed above. The contributions of our work are: 1. the integrate structure is proposed. 2. the power supply circuit of of core router and cellular BS is further ameliorated for the transition between sleep and working mode aforementioned. 3. based on these, a new mechanism is proposed for communications. The structure of this paper is organized as follows: Section II is proposed system model. The improved model is discussed in detail with C-RAN and ICN in this section.  In section III, The EE of this proposal is analyzed for both wired and wireless section of this proposal. The performance of circuit is analyzed and the transition scenario is raised in section IV. And the integrate working mechanism is demonstrated here. In section V, the proposal is compared with sleeping technology that raised previously. In section VI, we conclude this paper.

\section{Proposed System Model with C-RAN Architecture}

In C-RAN, typically, it consists of a cloud core network connecting with virtual BS pool, a high-speed switcher, Gbs/s or even Tbs/s fibre link, and several BSs that connecting with the switcher. While comparing with traditional radio access network, the cooperative multi-point (CoMP) accessing procedure can be more easily accomplished within the BS cluster with advantage that all transmission and receiving procedures are processing by virtual BS pool. As power of the BSs in a BS cluster are supplied by a power system uniformly, more energy can be saved because the number of air conditioner and other equipments can be greatly reduced by this architecture as talked aforementioned.
Within the massive MIMO background towards 5G wireless communications, in this paper, we separate the virtual BS pool into the plant and control center, switcher that connecting with cloud domain network. Which is shown as Fig. 1. The advantages are: the plant and control center provides power to each antenna of BS in the BS cluster; and it can decide to turn on or off the corresponding antenna with VLSI circuit and cable line that connecting with each of them. Here the antenna selection and radio frequency (RF) chain shortage problem is left for further study. In this paper, we say that the selected antenna or antenna arrays with corresponding UEs are already the optimal one. And we assume that there are enough RF chain for transmission with perfect channel state information (CSI).

\begin{figure}[!t]
\centering
\includegraphics[width=2.5in]{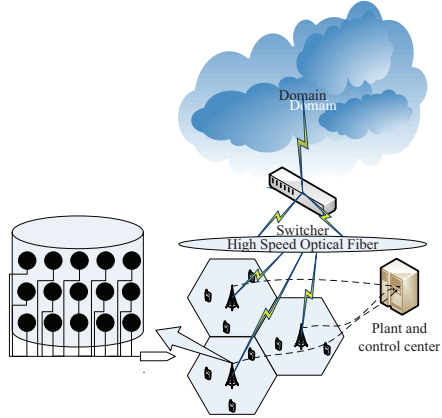}
\caption{Proposed System Model with C-RAN Architecture}
\label{fig_sim}
\end{figure}

\par
Another, the author in\cite{13} estimates that in core networks, for instance, with the rate of 100Mb/s, the terminal and switcher sections consume 30\% and the fiber links are responsible for 4.5\% of total energy consumption. Another conceive of this paper is, under the circumstances of C-RAN, it will be worthwhile and consequential to add the information centric network (ICN)\cite{12} leverage into this architecture as a feasible element with the purpose of alleviating the load burden of core network, saving more energy. The purpose is achieved by efficient distribution of information platform with caches that equipped with each cluster. It can provide services like point to point (P2P) in dedicate networks. The conception of ICN mainly relays on named data objectives (NDO) that independent distributed and stored in diverse locations. With NDO, identify and transmission processes of information are carried out by its name, thus its identify regardless of the location or transmission procedure. The transmission period can be depicted as Fig. 2.
\begin{figure}[!t]
\centering
\includegraphics[width=2.5in]{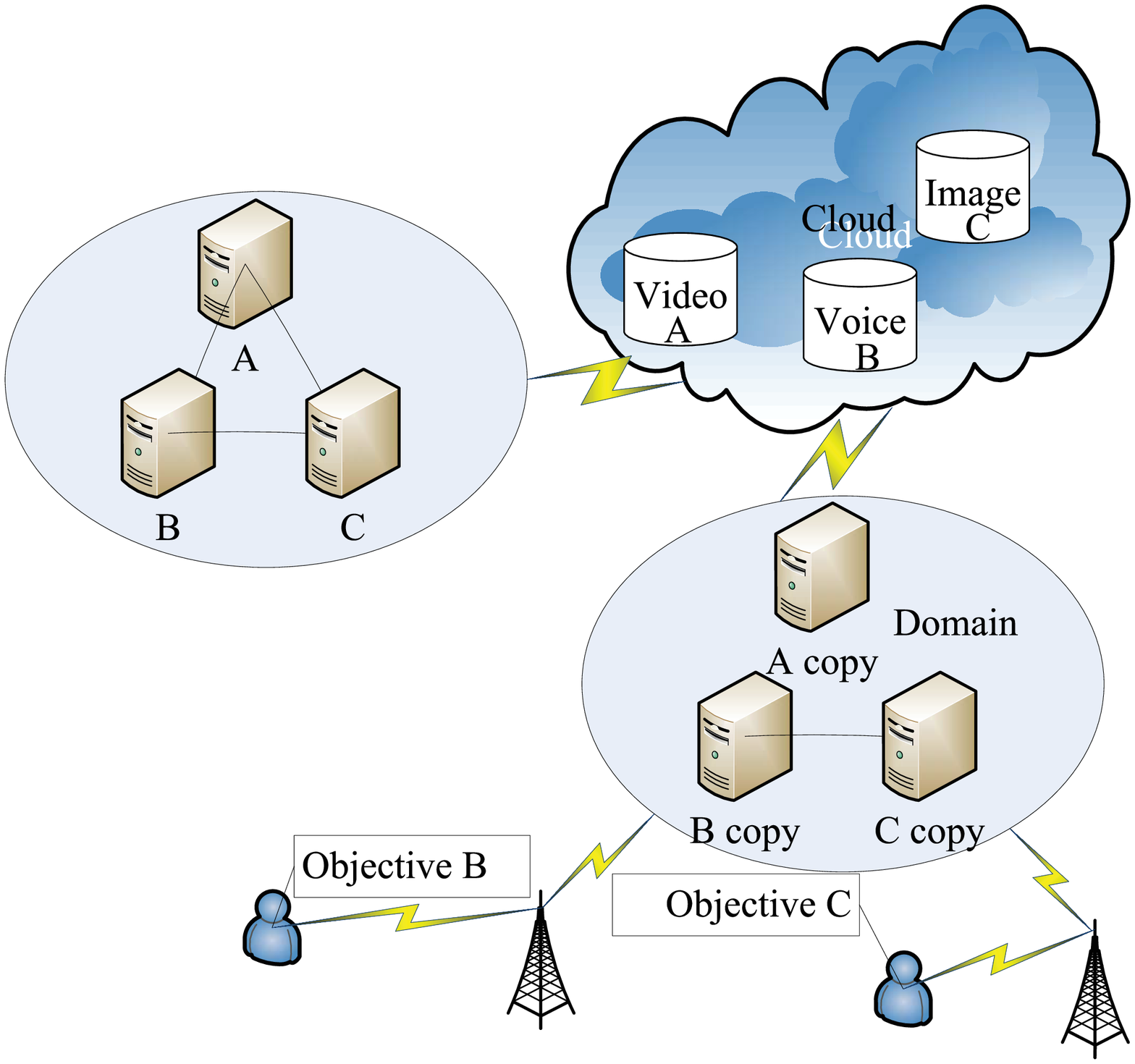}
\caption{the ICN based communication procedure}
\label{fig_sim}
\end{figure}

\par
The different is: in traditional IP protocol based core network, core devices such as switcher and router, will deliver the request by IP address. In contrast, with the help of caches that equipped within each core cluster, request will be packed and routed by NDO information no matter where the information located. The benefit relays on the disruption tolerance characteristic it is. And it is a more safer model which different with trust verification between client and server that traditional IP protocol based on but provides security transmission with name integrity and origin verification based on NDOs. Similar with P2P technology, ICN possesses the mobility and multihoming of core devices especially on condition that end to end connections. In addition, with caches equipped with each distribution cluster, it can be an energy efficient technology comparing with IP protocol,this issue can be proved in the following part of this work.
\section{Analysis of Energy Efficiency}
Typically, EE is defined as bit rate divided by energy that consumed to maintain the rate, thus EE formula with bit/joule can be expressed as:
\begin{equation}
\eta = R/P
\end{equation}
where $R, P$ denote for the bite rate and energy consumption that needed to feed the rate. While talking about EE optimization, we can have the formulas as:
\begin{equation}
 \begin{aligned}
& \max \eta\\
& s.t. ~~~\sum P \leq P_{sum},~~~P \in [0,P_{sum}]
 \end{aligned}
\end{equation}
where $P, P_{sum}$ are the required energy for transmission, the total power of BS in cellular or working router in core network.
\subsection{Energy Consumption of Wired section}
In this proposed regime, wired optical fiber between BS and cluster added with optical fiber between different clusters in core networks consist of the wired transmission line. Typically, in wired networks, energy consumption mainly derive from the consumption of transmission signal, core routers (optical cross connect with abbreviation of OXC and transponders), inline optical fiber together with amplifiers established in. While working, the sleep mode is further adopted for OXC, transponder together with the optical fiber for energy efficiency as prior literatures. And we assume that the router is optimal throughout the cloud network. The power consumption model can be given by:
\begin{equation}
P_o = P_s + nP_c
\end{equation}
where $P_o, P_s, P_c, n$ stand for the energy consumption of total wired section, transmission signal, core routers and the number of it. Under the assumption that optical fiber is divided by fiber section with Erbium Doped Fibre Amplifier (EDFA) and section without it and number of core router, denote $\alpha, L, L_{edfa}, l$ for the number of EDFA that equipped within the optical fiber, distance from required information to the BS, per length that EDFA can back up, remained distance plus the distance from cluster to BS, $P_s$ can be evaluated by:
\begin{equation}
P_s = P_t +\alpha P_{edfa}
\end{equation}
where $\alpha$ can be indicated as:
\begin{equation}
\alpha  = \lceil \frac{L}{L_{edfa}} \rceil -1
\end{equation}

Typically the distance each EDFA can compensate is 80km. On condition that the gain of optical fiber without EDFA is $G$, attenuation of optical fiber $\beta$, thus the section of optical fiber that without compensated by the EDFA can be formulated by:
\begin{equation}
l = L- \alpha L_{edfa}+L_{BS}
\end{equation}
where $L_{BS}$ stands for the distance from cluster to BS. That is because of the regime that by sleep mode of non-active OXC and transponder, from cluster to BS, there are same number of transponders and optical fibers for each active BS. In this case, $P_{ase}$ can be expressed by:
\begin{equation}
P_{ase} = Gle^{-\beta l}\frac{P_t^{n-1}}{16*40}n_g
\end{equation}
where $P_t^{n-1}$ stands for the power that needed in the former hop, which is of the same power value at the last end of optical fiber equipped with EDFA on account of the hypothesis that EDFA can compensate all of the losses during transmitting. And $n_g$ stands for the transmission bit rate by $Gbps$. Assume that Cisco CRS-1 with 16 slots series is used as core router, which will consume about 10.9KW power for full-duplex mode, then while transmitting, $P_t^{n-1}$ can be expressed by $P_t^{n-1} = 10.9KW/16$ for each working slot on condition that transition between sleep mode and working mode added. Finally, $P_t$ can be expressed by:
\begin{equation}
P_t = \frac{P_t^{n-1}}{16*40}n_g+p_{ase}
\end{equation}

\subsection{Energy Consumption of Wireless section}
In Massive MIMO with cylinder disposition, typically, there is one massive MIMO BS with hundreds of antennas (here the number denotes as N) and K UEs where $K<<N$ (typically, for a massive MIMO system, there are 100$\sim$200 antennas distributed by cylinder shape that serves 42 UEs.). For simplicity, we assume that the ergodic capacity can be formulated by perfect channel state information (CSI) through pilot of the transmitter.ZF-BF is used for interference eliminating where the ZF-BF matrix is $W = [w_1,w_2,...,w_k]$ . In ZF-BF, take $W$ as the beamformer for $ w_k$ channels within a $N \times K$ matrix, where $ w_k$ a $N\times 1$ vector. Here $h_k$ , channel from BS to UE is $1 \times N $ complex-Gaussian entries vector with zero mean and unit variance. Then the total channel matrix, a $ K \times N $ complex matrix, can be defined as $ H = [h_{1}^{T}, h_{2}^{T},..., h_{k}^{T}]$ . Then the ZF-BF matrix can be expressed as:
\begin{equation}
W = H^{T}(H H^{T})^{-1}
\end{equation}
where $ (.)^H $ and $ (.)^{-1}$ stand for conjugate and inverse transpose of a matrix separately. Then the received signal of the \textit{k}th user can be expressed as:
\begin{equation}
y = h_{k} \frac{w_{k}}{\sqrt{\gamma}}x_{k} + h_{k} \sum_{i=1,i \neq k}^{K}\frac{w_{i}}{\sqrt{\gamma}}x_{i} +n_{k}
\end{equation}
where $x_{k}$ denotes transmitted signal at BS terminal of the \textit{k}th user, which can be expressed as $x_{k}= \sqrt{P_{k}}s_k$  with $P_k$, transmission power of UE \textit{k}. And $s_k $ is transmission signal at BS terminal of \textit{k}th user that obeys complex Gaussian distribution with zero mean and uniform variance,  $n_k$ noise signal with zero mean and  $N_{0}$ variance, $\gamma$  the normalization factor of signal of the \textit{k}th user with expression $ \gamma = \lVert W \rVert_{F}^{2} / K $ . And $ \lVert . \rVert_{F}^{2} $ denotes for matrix Frobenius norm.

\par
Suppose channels are constituted by components carriers (CCs) where CCs are sufficient for usage of kth users, and maximum bandwidth each CC can carry is $B_{CCs}$. If $ SINR_k $  is taken as signal to interference plus noise ratio of the \textit{k}th user. Received bit rate of \textit{k}th UE can be expressed as:
\begin{equation}
R_k = B_{CCs} \log_2 (1+SINR_k)
\end{equation}
as the assumption, $B_{CCS}$ is known with a constant value, $ SINR_k $  should be focused. For simplicity, we assume that each UE associated with two antennas while transmitting, thus suppose the power of antenna array with number of $n$ in each BS is $P_{bs}$, while turn on the associate antennas with and turn of the unexpected antennas to sleep, for \textit{k}th UE, the power is allocated by $ \beta = P_{bs}2K/n $ . Taking $\rho = \beta / N_0$ as the signal to noise (SNR), Thus, the SINR of UE k can be formulated as:
\begin{equation}
SINR_k = \frac{\rho|h_k w_k|^2} {\rho \sum_{j=1,j \neq k}^{j=K}|h_k w_j|^2+1}
\end{equation}
\par
According to equ.9, the beamforming vector can be deduced as:
\begin{equation}
w_k = \frac{h_{k}^{H}}{\|h_k\|}, ~~k = 1,...,K
\end{equation}
as described aforementioned, ZFBF can eliminate the intra-interference. 
Achievable ergodic rate of \textit{k}th UE can be described as:
\begin{equation}
\begin{aligned}
{R_k}' &= E\{B_{CCs} \log_2 (1+\frac{\rho K}{tr\{(H H^H)^{-1}\}})\}\\
       &\approx B_{CCs} \log_2 (1+\frac{\rho K}{E\{tr\{(H H^H)^{-1}\}\}})
\end{aligned}
\end{equation}

\par
According to random matrix theory, the matrix $HH^H$ is a central Wishart matrix. suppose the number of antenna is larger than UE, then the matrix would be a $K \times K$ central Wishart matrix with N degrees of freedom, and its covariance matrix $\textbf{I}$, a unitary matrix.  Then $tr\{(H H^H)^{-1}\}$ can be deduced as:
\begin{equation}
\begin{aligned}
tr\{(H H^H)^{-1}\} &= tr\{(H H^H)^{-1}HH^H(H H^H)^{-1}\}\\
                   &=tr\{W^HW\} =\|W\|_{F}^2
\end{aligned}
\end{equation}
as K, M(associate antennas) growing with a constant ratio $\alpha = M/K $, it converges to a fixed deterministic value:$E\{\|W\|_{F}^2\} = K/(M-K)$, then max-sum ergodic of per cell can be formulated as:
\begin{equation}
\begin{aligned}
R_{sum} = B_{CCs} \log_2 (1+\frac{\rho K}{E\{\|W\|_{F}^2\}})
\end{aligned}
\end{equation}
As estimated aforementioned by CMCC, the machine room is responsible for 28\% energy consumption, which is ignored in most of prior
literatures. Thus with energy consumption of machine room $P_{mr}$ added, the energy consumption can be described by:
\begin{equation}
P_{sum}= P_{BS}+P_{mr}
\end{equation}
\section{Proposed Transition Smooth Scheme}
\subsection{Analysis of transition period}
As surge pulse (thunder, pulse surge, etc) can also bring serious damage to electron devices, according to the circuit theory, typically, two bypass capacitances are used for eliminating the surge or high frequency current to protect the load, thus the antenna in this paper. The protection circuit model is shown in Fig.3. Tanking the antenna as one capacitance devices with value of $C3$. On condition that it can be easily damaged by surge pulse current, which is similar to the performance of capacitance. The capacitance value of the whole circuit can be expressed with $C = 1 / C1 +1 / C2 +1 / C3 $. According to circuit theory, we can adopt RLC circuit model to describe the power supply system for transition between on-off period of antennas in cellular system. Which is shown as in Fig. 4.
\begin{figure}[!t]
\centering
\includegraphics[width=2.5in]{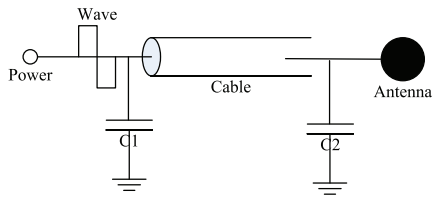}
\caption{the protection circuit}
\label{fig_sim}
\end{figure}

\begin{figure}[!t]
\centering
\includegraphics[width=2.5in]{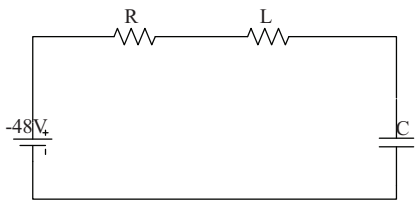}
\caption{Simplified RLC model of power system}
\label{fig_sim}
\end{figure}

On condition that power supply is turned off, according to the source free series RLC circuit model, while applying the KLV around the loop, we can have:
\begin{equation}
 \begin{aligned}
\frac{d^2i}{dt^2} + \frac{R}{L}\frac{di}{dt} + \frac{i}{LC} = 0
 \end{aligned}
\end{equation}
where $v(0) = \frac{1}{C}\int_{-\infty}^{0}idt=V_0, i = C\frac{du_c}{dt}$, $i(0)=I_0$. By solving the equation ,we can have:
\begin{equation}
 \begin{aligned}
i = &Ae^{st} \\
s = &-\alpha \pm \sqrt{\alpha^2-\omega^2}
 \end{aligned}
\end{equation}
where $\alpha = \frac{R}{2L}$ is the neper frequency and $\omega = \frac{1}{\sqrt{LC}}$ is the resonant frequency. Then the three types of solutions can be defined as: $\alpha > \omega$ is the overdamped case, $\alpha = \omega$ is the critically damped case, $\alpha < \omega$ is the underdamped case. In electron devices, typically, the impedance of resistance can be taken as zero. And according to wsitching law $u_{c}(0^+)=u_{c}(0^-)=U_0, i(0^+)=i(0^-)=0$, the voltage of capacitance will be:
 \begin{equation}
 \begin{aligned}
u_c(t) = Ucos(\omega t)
 \end{aligned}
\end{equation}
While adopting the expressions $i = C\frac{du}{dt}, u_l = L\frac{di}{dt}$, we can have the voltage of inductance by several operations:
\begin{equation}
u_l(t) = - Ucos(\omega t)
\end{equation}
 \par
 Similarly, while $u_c, U$ denotes for the capacitance voltage and power source voltage respectively, on the power on moment, applying the series RLC step response theory and after a series of operations, we can have:
\begin{equation}
 \begin{aligned}
 LC\frac{d^2u}{dt^2} + RC\frac{du}{dt} + u_c = U
 \end{aligned}
\end{equation}
while ignoring the circuit resistance,  $u_c(t)$ can be further changed to:
\begin{equation}
 u_c(t) = U(1-\cos\omega t)
\end{equation}
then after adopting the functions $i = C\frac{du}{dt}, u_l = L\frac{di}{dt}$, the voltage of inductance $u_l(t)$ can be expressed as:
\begin{equation}
u_l(t) = Ucos(\omega t)
\end{equation}

\subsection{Proposed Scheme towards on-off Transition}
Typically, while analysing the second order oscillation circuit, the $\tau = 2\pi \sqrt{LC}$ is used to denote the oscillation period with $\omega_c = 2\pi / \tau$. Denote $\tau$ as working time of source free period, and one $T$ denotes for one period of on-off transition. Then the voltage function of $u_c(t)$ can be combined as :
\begin{equation}
 u_c(t)=
\begin{cases}
 U \cos(2 \pi(\frac{t}{\tau})) , \textup{source free period}\\
 U(1-\cos (2 \pi(\frac{t}{\tau}))) , \textup{step response period}
\end{cases}
\end{equation}
similarly,the function of $u_l(t)$ can be expressed as:
\begin{equation}
 u_l(t)=
\begin{cases}
 -U \cos(2 \pi(\frac{t}{\tau}) ) ,  \textup{source free period}\\
 U\cos (2 \pi(\frac{t}{\tau})) , \textup{step response period}
\end{cases}
\end{equation}

It is known that the voltage supply is some kind of rectangle wave.  It can be further took as two rectangle wave with different positive and negative values during the source free and step respond period. 
While talking about the protection of working life electronic device, typically, during one on-off period,the working voltage should not exceed the nominal voltage. In this case, the expression of protection of electron devices working life can be written as:
\begin{equation}
\begin{aligned}
u_c(t) &= Ucos(2 \pi(\frac{t_1}{\tau}))+Ucos(2 \pi(\frac{t_2}{\tau}))-U \leq U \\
u_l(t) &= -Ucos(2 \pi(\frac{t_1}{\tau}))-Ucos(2 \pi(\frac{t_2}{\tau})) \leq U
\end{aligned}
\end{equation}
by solving the problem, the expression of $t_1, t_2$ should be further devided by $t_1 \leq t_2$, $t_1 > t_2$. Considering that the solutions of these two conditions are some kind of similarity, the solutions are given here with expression by:
\begin{equation}
t_2 \leq \frac{\tau arccos(\frac{1}{2}-cos(2\pi (\frac{t_1}{\tau})))}{2\pi}
\end{equation}
or
\begin{equation}
t_1 < \frac{\tau arccos(\frac{1}{2}-cos(2\pi (\frac{t_2}{\tau})))}{2\pi}
\end{equation}

\par
The processing procedure can be accomplished with some pseudo-algorithm while communicating. And in this paper we adopt a simple antenna and RF chain selection mechanism with  $I_{a}, I_{rf}, S_{a}, S_{rf}, IN_r, IN_c, IN_f$ stand for idle antenna, idle RF, state of the selected antenna and state of selected RF, required information by the request, information located in the cluster caches and information in far away clusters respectively. The antenna and RF chain numbers are assumed sufficient enough for each request. Perfect CSI feature is adopted simultaneously. For processing procedure, the pseudo-algorithm is described as in Algorithm 1. 

\begin{algorithm}[!ht]
\caption{Processing Angorithm}
\begin{algorithmic}[1]
\STATE active UE comes into the BS cluster
 \STATE Search for the nearest BS, selection antenna and RF-chain with signal strong link mechanism
 \IF{$I_{a}>0;  I_{rf}>0$}{
     \IF{$S(a)=ON; S(rf)=ON$}{
         \IF {$IN_r = IN_c$}
              \STATE $IN_c \to IN_r$;~estimate $t_2$ with $t \mod \tau /4$;
              \STATE delay $ t_1,$ set $S(a)=OFF, S(rf)=OFF$
         \ELSE [$IN_r \neq IN_c$]
                \STATE Flood request with NDO;
                \STATE $IN_f \to IN_r$estimate $t_2$;
              \STATE delay $ t_1,$ set $S(a)=OFF, S(rf)=OFF$
         \ENDIF
     }
     \ELSE[$S(a)\neq ON || S(rf)\neq ON$]
            \STATE wait and listen
            \IF{$S(a)=ON; S(rf)=ON$}
            \STATE goto step 4
            \ELSE
            \STATE goto step 14
            \ENDIF
     \ENDIF
 }
 \ELSE[$I_{a}=0  ||  I_{rf}=0$]
 \STATE wait and listen
        \IF{$I_{a}>0;  I_{rf}>0$}
        \STATE goto step 3
        \ELSE
        \STATE goto step 22
        \ENDIF
 \ENDIF
\end{algorithmic}
\end{algorithm}

\section{Simulation Results}
In this section, we evaluate our proposed scheme and compare with the state-of-art techniques in terms of EE performance. General simulation elements are shown in Table I. We evaluate the performance with EE feature of our proposed scheme while comparing with the latest EE proposal with sleep mode both in wired and wireless section.
\par
Suppose the modulation mode is 16 QAM in wireless communications, one cluster associates with 20 BS, we can get the EE performance at wireless section as Fig. 5. It is obviously that the proposed regime with C-RAN possesses better EE performance because the energy consumption of machine is greatly deduced comparing with traditional architecture, where each BS equipped with one machine room. And in the curve of traditional sleep scheme, each sawtooth denotes for a machine room been added on condition that the number of UEs exceeding maximum UEs' number one BS can bear. And while the number of UEs growing, the EE curves trend to flat, that is because of the total maximum UEs' number restrict of one cluster.
\begin{table}[!t]
\renewcommand{\arraystretch}{1.3}
\caption{Simulation Elements}
\label{table_example}
\centering
\begin{tabular}{l l } 
\hline\hline 
Parameters & Value  \\ [0.5ex] 
\hline 
40Gbps port & 400W  \\ 
EDFA  & 4W/Gps \\
OXC with degree $d_f$ & 150W+135$d_f$W \\
Gain of fiber & 0.99 \\
Fiber attenuation Coefficient & 0.3dB/km \\
Machine room & 480W \\
Number of antennas M &200 \\
wireless peak bit rate & 200MHz\\
Bandwidth $B_{CCs}$ &5MHz \\[1ex] 
\hline 
\end{tabular}
\label{table:nonlin} 
\end{table}
\par
In the wired section, we assume that the distance between two hop are 100km, in addition, the distance from BS to cluster is 10km that connected by optical fiber. The EE performance in terms of transponder and port sleep mode with ICN are shown as Fig. 6. We take the number of hops standing for the location of required information from UEs in BS coverage area. When required information located in ICN cache of the cluster servicing the BSs, that will be one hop, otherwise, number of hops added by location. In this case, as described in the figure, EE performance declines with number of hops adding. And the curvature change around 40 in Fig. 6 manifests that more transponders are added for transmission because of the maximum 40Gbps one optical fiber can sustain. It is clearly that better EE performance can be obtain with ICN technology in wired core while serving as the data center of wireless communication as descibed in Fig. 6. That is because that with ICN, the UE can always get information with least routing distance. And with the help of sleep mode in unexpected router and optical fiber, great deal of energy can be saved.
\begin{figure}[!t]
\centering
\includegraphics[width=3in]{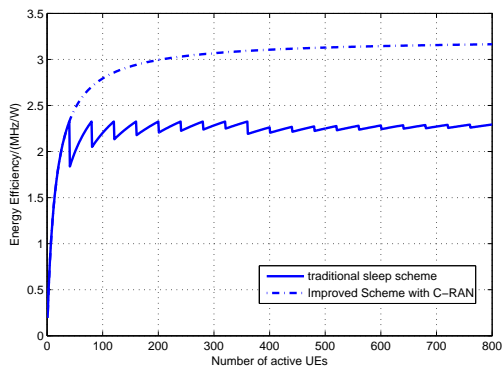}
\caption{Energy Efficiency Performance of Wireless Section With Improved C-RAN}
\label{fig_sim}
\end{figure}
\begin{figure}[!t]
\centering
\includegraphics[width=3in]{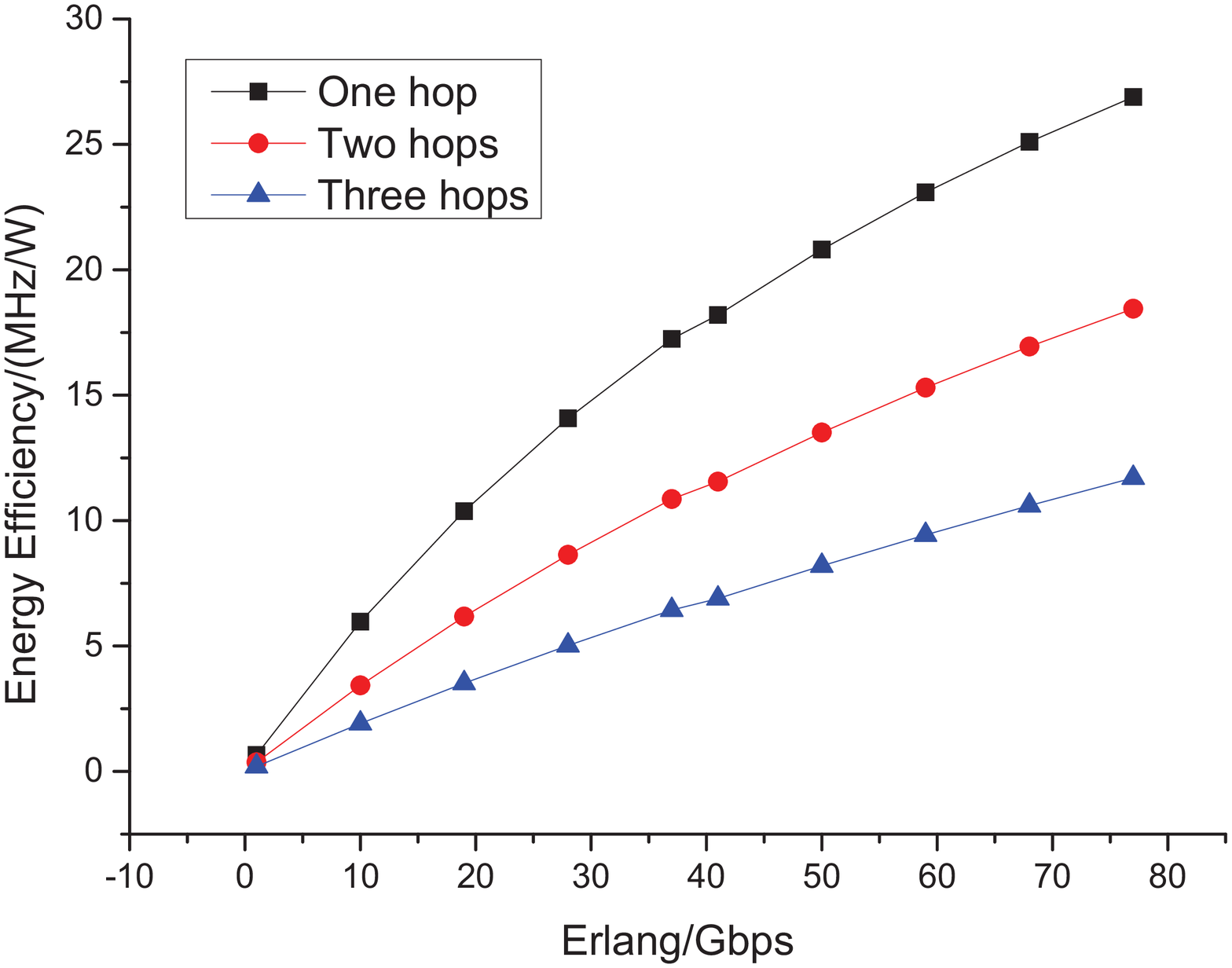}
\caption{Energy Efficiency Performance of Core Network with ICN Technology}
\label{fig_sim}
\end{figure}
\section{Conclusion}
In this paper, one modified ICT architecture is proposed based on the C-RAN and ICT technologies towards EE. Detail construction of the improved policy is described and the EE analysis is demonstrated. Based on all of these, the decision scenario of sleep scheme for EE is proposed with transition between sleep and working mode both in wireless and wired section. Sleep policy can be more realistic with the help of this proposed method. The performance of this regime is compared with simple sleep scenarios that proposed latest for EE. And it is shown that the policy of this paper possess better EE feature. This work can be further combined with imperfect CSI and antenna selection method for more reality circumstance.






\begin{thebibliography}{10}
\providecommand{\url}[1]{#1}
\csname url@samestyle\endcsname
\providecommand{\newblock}{\relax}
\providecommand{\bibinfo}[2]{#2}
\providecommand{\BIBentrySTDinterwordspacing}{\spaceskip=0pt\relax}
\providecommand{\BIBentryALTinterwordstretchfactor}{4}
\providecommand{\BIBentryALTinterwordspacing}{\spaceskip=\fontdimen2\font plus
\BIBentryALTinterwordstretchfactor\fontdimen3\font minus
  \fontdimen4\font\relax}
\providecommand{\BIBforeignlanguage}[2]{{%
\expandafter\ifx\csname l@#1\endcsname\relax
\typeout{** WARNING: IEEEtran.bst: No hyphenation pattern has been}%
\typeout{** loaded for the language `#1'. Using the pattern for}%
\typeout{** the default language instead.}%
\else
\language=\csname l@#1\endcsname
\fi
#2}}
\providecommand{\BIBdecl}{\relax}
\BIBdecl

\bibitem{1}
H.~Chen, Y.~Jiang, J.~Xu, and H.~Hu, ``Energy-efficient coordinated scheduling
  mechanism for cellular communication systems with multiple component
  carriers,'' \emph{IEEE J. Sel. Areas Commun.}, vol.~31, no.~5, pp. 959--968,
  May. 2013.

\bibitem{2}
H.~Li, L.~Song, and M.~Debbah, ``Energy efficiency of large-scale multiple
  antenna systems with transmit antenna selection,'' \emph{IEEE Trans.
  Commun.}, vol.~62, no.~2, pp. 638--647, Feb. 2014.

\bibitem{5}
Z.~Niu, Y.~Wu, J.~Gong, and Z.~Yang, ``Cell zooming for cost efficient green
  cellular networks,'' \emph{IEEE Commun. Mag.}, vol.~48, no.~11, pp. 74--79,
  Nov. 2010.

\bibitem{6}
Q.~Zhang, C.~Yang, H.~Haas, and J.~Thompson, ``Energy efficient downlink
  cooperative transmission with bs and antenna switching off,'' \emph{IEEE
  Trans. Wireless Commun.}, vol.~PP, no.~99, pp. 1--14, May. 2014.

\bibitem{7}
X.~Zhang, S.~Zhou, Z.~Niu, and X.~Lin, ``An energy-efficient user scheduling
  scheme for multiuser mimo systems with rf chain sleeping,'' in \emph{IEEE
  WCNC}, Apr. 2013, pp. 169--174.

\bibitem{8}
M.~Ju, H.-K. Song, and I.-M. Kim, ``Joint relay-and-antenna selection in
  multi-antenna relay networks,'' \emph{IEEE Trans. Commun.}, vol.~58, no.~12,
  pp. 3417--3422, Dec. 2010.

\bibitem{9}
Z.~Zhou, S.~Zhou, J.~Gong, and Z.~Niu, ``Energy-efficient antenna selection and
  power allocation for large-scale multiple antenna systems with hybrid energy
  supply,'' in \emph{IEEE Globecom Accepted}, Austin, TX, USA, Dec. 2014.

\bibitem{11}
Z.~Kong, J.~Gong, C.-Z. Xu, K.~Wang, and J.~Rao, ``ebase: A baseband unit
  cluster testbed to improve energy-efficiency for cloud radio access
  network,'' in \emph{IEEE ICC}, Jun. 2013, pp. 4222--4227.

\bibitem{16}
A.~Dhaini, P.-H. Ho, G.~Shen, and B.~Shihada, ``Energy efficiency in tdma-based
  next-generation passive optical access networks,'' \emph{IEEE/ACM Trans.
  Netw.}, vol.~22, no.~3, pp. 850--863, Jun. 2014.

\bibitem{15}
A.~Rasmussen, M.~Yankov, M.~Berger, K.~Larsen, and S.~Ruepp, ``Improved energy
  efficiency for optical transport networks by elastic forward error
  correction,'' \emph{IEEE/OSA J. Opt. Commun. and Netw.}, vol.~6, no.~4, pp.
  397--407, Apr. 2014.

\bibitem{13}
A.~Coiro, M.~Listanti, A.~Valenti, and F.~Matera, ``Reducing power consumption
  in wavelength routed networks by selective switch off of optical links,''
  \emph{IEEE J. Sel. Top. Quant.}, vol.~17, no.~2, pp. 428--436, March 2011.

\bibitem{12}
B.~Ahlgren, C.~Dannewitz, C.~Imbrenda, D.~Kutscher, and B.~Ohlman, ``A survey
  of information-centric networking,'' \emph{IEEE Commun. Mag.}, vol.~50,
  no.~7, pp. 26--36, July 2012.

\end{thebibliography}
%
\begin{bibliographystyle}{IEEEtran}
\begin{bibliography}{IEEEabrv,secondpaper}
\end{bibliography}
\end{bibliographystyle}

\end{document}